# Mapping Intrinsic Electromechanical Responses at the Nanoscale via Sequential Excitation Scanning Probe Microscopy Empowered by Deep Data

Boyuan Huang [1,*], Ehsan Nasr Esfahani [1,*], and Jiangyu Li [1,2,†]

1. Department of Mechanical Engineering, University of Washington, Seattle, WA 98195-2600, USA
2. Shenzhen Key Laboratory of Nanobiomechanics, Shenzhen Institutes of Advanced Technology, Chinese Academy of Sciences, Shenzhen 518055, Guangdong, China

## Abstract

Ever increasing hardware capabilities and computation powers have made acquisition and analysis of big scientific data at the nanoscale routine, though much of the data acquired often turns out to be redundant, noisy, and/or irrelevant to the problems of interests, and it remains nontrivial to draw clear mechanistic insights from pure data analytics. In this work, we use scanning probe microscopy (SPM) as an example to demonstrate deep data methodology, transitioning from brute force analytics such as data mining, correlation analysis, and unsupervised classification to informed and/or targeted causative data analytics built on sound physical understanding. Three key ingredients of such deep data analytics are presented. A sequential excitation scanning probe microscopy (SE-SPM) technique is first adopted to acquire high quality, efficient, and physically relevant data, which can be easily implemented on any standard atomic force microscope (AFM). Brute force physical analysis is then carried out using simple harmonic oscillator (SHO) model, enabling us to derive intrinsic electromechanical coupling of interests. Finally, principal component analysis (PCA) is carried out, which not only speeds up the analysis by four orders of magnitude, but also allows a clear physical interpretation of its modes in combination with SHO analysis. A rough piezoelectric material has been probed using such strategy, enabling us to map its intrinsic electromechanical properties at the nanoscale with high fidelity, where conventional methods fail. The SE in combination with deep data methodology can be easily adapted for other SPM techniques to probe a wide range of functional phenomena at the nanoscale.

---

[*] These authors contributed equally to the work.

[†] Author to whom the correspondence should be addressed to.

## Conceptual Insights

Mapping weak electromechanical coupling quantitatively at the nanoscale using scanning probe microscopy (SPM) is very challenging. While conventional resonance enhanced techniques can be used to improve sensitivity, the measurements are prone to crosstalks with topography, elasticity, and dissipation. In this work, we report sequential excitation (SE) that excites the sample using a sequence of AC waveforms with distinct frequencies, each capturing cantilever-sample resonance at selected spatial points that are unique. In a sense, such SE strategy is analogous to super-resolution microscopy in biology that turns specific fluorescent molecules on and off in a sequential manner for imaging, after which the sequence of data can be reconstructed to determine the intrinsic electromechanical response using either physical based simple harmonic oscillator (SHO) model or multivariate statistical tools such as principal component analysis (PCA). Of particular interest is a clear mechanistic understanding of the unsupervised PCA acquired through sound physical modeling, making it possible to speed up physical analysis by at least four orders of magnitude. This work provides us an ideal playground for targeted and causative deep data methodology, illustrating three key ingredients including: (1) developing innovative experimental and/or computational methodologies to acquire high quality (less noisy), efficient (less redundant), and physically relevant scientific data to enable deep analysis; (2) learning clear mechanistic insights from the data, guided by the underlying physical principles; and (3) accelerating and enhancing physical understanding by informed and targeted big data analytics.

**Introduction**

The fusion of scientific research and big data has provided unprecedented opportunity for accelerated discovery, understanding, and innovation [1–6], yet it also imposes new challenges for scientists to adjust, adapt, and thrive in the face of daunting data volume [7,8]. Ever increasing hardware capabilities and computation powers have made acquisition and analysis of big scientific data routine, though much of the data acquired often turns out to be redundant, noisy, and/or irrelevant to the problems of interests, and it remains nontrivial to draw clear mechanistic insights from brute force data analytics [2,7,9]. As such, there is a strong desire to push big data toward deep data, *i.e.*, from data mining, correlation analysis, and unsupervised classification to causative data analytics that fuse scientific knowledge of physics, chemistry, and biology into big data [2,3,5,6,9,10], and thus from brute forces to informed and/or targeted strategies. We believe that three key questions need to be answered to enable such vision are: (1) how do we develop innovative experimental and/or computational methodologies to acquire high quality (less noisy), efficient (less redundant), and physically relevant scientific data to enable deep analysis; (2) how can we learn clear mechanistic insights from the data, guided by the underlying physical principles; and (3) how can we accelerate and enhance physical understanding by informed and targeted big data analytics. Scanning probe microscopy (SPM) is capable of acquiring multi-dimensional physical datasets in the form of high resolution images and spectroscopy [11,12], providing us an ideal playground for deep data methodology. In this work, we demonstrate the essence of such deep data analysis via a simple yet revealing case study that incorporates all the three key ingredients above, enabling us to resolve a long-standing challenge in scanning probe microscopy - mapping weak intrinsic response at the nanoscale quantitatively [13–15]. Of particular interest is a clear mechanistic understanding of the unsupervised principle component analysis (PCA) [16–18] acquired through sound physical principle, which makes it possible to speed up physical analysis by at least four orders of magnitude.

**Results and Discussion**

*Sequential excitation*

As the first step, we develop a sequential excitation scanning probe microscopy (SE-SPM) technique to acquire high quality [19], efficient, and physically relevant data in frequency domain. The method can be easily implemented in any standard atomic force microscope (AFM) without the need for any additional hardware and instrumentation. The code used for SE-PFM analysis can be accessed at online [20]. To this end we note that majority of SPM measurements deduce physical properties of samples from the interactions between a cantilever with a sharp tip and sample surface, as schematically shown in Fig. 1, and the dynamics of the interaction can be described well by a simple harmonic oscillator (SHO) model [21,22],

$$A(\omega) = \frac{A_0 \omega_0^2}{\sqrt{(\omega_0^2 - \omega^2)^2 + (\frac{\omega_0 \omega}{Q})^2}}, \quad (1.1) \qquad \phi(\omega) = \tan^{-1}[\frac{\omega_0 \omega}{Q(\omega_0^2 - \omega^2)}], \quad (1.2)$$

where $A_0$, $\omega_0$, and $Q$ are intrinsic electromechanical response (piezoelectricity), resonant frequency (elasticity), and quality factor (energy dissipation) of the system that we are interested in learning, while $\omega$ is the excitation frequency with $A(\omega)$ and $\phi(\omega)$ as the corresponding amplitude and phase that are directly measured in experiment. As such, it is desirable to acquire the amplitude $A(\omega, x, y)$ and phase $\phi(\omega, x, y)$ over a two-dimensional (2D) space $(x, y)$ of sample surface as well as a frequency spectrum of excitation ($\omega$), from which $A_0$, $\omega_0$, and $Q$ can be solved from Eq. (1). Conventional techniques such as dual amplitude resonance tracking (DART) [22,23] and band excitation (BE) [24,25] synthesize AC waveform combining either two distinct or a band of frequencies to excite the sample (Fig. 1). The former yields only two set of data, not amenable for reliable analysis, and the resonance tracking is not always robust, especially for rough sample surfaces. The latter, on the other hand, distributes excitation energy over a frequency band, reducing signal strength substantially and thus acquired data are often noisy. Newly developed general mode (G-mode) SPM records complete time- instead of frequency-domain data [7,26], yet a substantial portion of data is either redundant or irrelevant for the problem of interests, and it requires sophisticated instrumentation that is not easily accessible. Therefore, there is still a strong desire for innovative yet simple and easily accessible approach to acquire high quality (less noisy), efficient (less redundant), and physically relevant SPM data to enable deep analysis.

To this end, we develop sequential excitation (SE) that excites the sample using a sequence of AC waveforms with distinct frequencies $\omega_j$, as shown in Fig. 1, wherein the excitation energy is concentrated on only one frequency at a time instead of being distributed over a band of spectrum, ensuring that the signal is strong and the response is not noisy. In such a setup, each excitation frequency captures cantilever-sample resonance at selected spatial points that are unique, ensuring that the data is relevant yet not redundant. Furthermore, no resonance tracking is needed as in DART, ensuring that the measurement is robust and reliable. In a sense, such strategy of SE is analogous to super-resolution microscopy in biology that turns specific fluorescent molecules on and off in a sequential manner for imaging [27,28], wherein we excite specific resonances of different points sequentially using distinct frequencies. Our approach, however, requires no extra hardware and further instrumentation in a standard AFM, and it can be easily implemented, making it widely accessible.

As a demonstration, we study piezoresponse force microscopy (PFM) of a PZT ceramic under SE, wherein a sequence of its amplitude mappings $A(\omega_j, x, y)$ are shown in Fig. 2, obtained using AC excitation frequencies ranging from 320 to 400 kHz. The drifting between different scans has been corrected as detailed in Supporting Information (SI). It is observed that the PFM amplitude is very sensitive to the excitation frequency $\omega_j$, as expected, and there are substantial amplitude changes when the excitation frequency varies. In addition, substantial spatial heterogeneity is observed within each mapping, reflecting possible variations in intrinsic piezoelectricity, elasticity, energy dissipation, or their combination. Such crosstalk makes it difficult to determine intrinsic SPM response quantitatively, and it is necessary to deconvolute these different effects. In fact, our work was originally motivated by this very issue, which has important implication in nanoscale probing of electromechanical coupling ubiquitous in nature that underpins functionalities of both synthetic materials and biology for information processing as well as energy conversion and storage [11,29–35]. While dynamic strain-based SPM techniques have emerged as a powerful tool to investigate electromechanical coupling at the nanoscale in the last decade [11], which are known as PFM for piezoelectrics and ferroelectrics [34,36–40] and as electrochemical strain microscopy (ESM) for electrochemical systems [41–46], determining intrinsic electromechanical response remains challenging due to its crosstalk with topography, elasticity, and energy dissipation. SE-PFM makes it possible to overcome such difficulties, though we must reconstruct data to determine the intrinsic response, analogous to super-resolution microscopy in biology. In this regard, the three-dimensional (3D) data sets of amplitude $A(\omega, x, y)$ and phase $\phi(\omega, x, y)$ obtained from SE are amenable to both physics-based SHO analysis and statistics based PCA, making deep data analysis possible.

*Physical analysis by simple harmonic oscillator model*

We start with brute force physical analysis accomplished by fitting 3D datasets of amplitude $A(\omega, x, y)$ at each pixel $(x, y)$ using SHO model represented by Eq. (1.1). This is demonstrated by one representative pixel in Fig. 3a, yielding a resonant frequency of 347.8kHz, quality factor of 33.23, and intrinsic electromechanical response of 15.25pm at that particular point. Note that the sample surface is rather rough as revealed by its topography (Fig. 3b), which imposes substantial difficulty for DART-PFM, yet such SHO analysis can be easily applied to each pixel of SE-PFM to reconstruct the mappings of intrinsic amplitude, resonant frequency and quality factor, as shown in Fig. 3c-e. Indeed, there is strong spatial variation in intrinsic amplitude mapping, though little correlation is seen between topography (Fig. 3b) and amplitude (Fig. 3c), even in regions with substantial roughness, for example in the valley on the top part of the mapping marked by the dotted red square. The mapping of $R^2$, a statistical measure known as the fitting coefficient of determination accessing how close the data are to the fitted regression line, is presented in Fig. S1 in SI, revealing values ranging from 0.85 to 0.99 and thus a high fidelity of SHO analysis. This demonstrates the capability of SE-PFM even for highly inhomogeneous and rough samples. Such capability, however, is beyond the conventional DART-PFM, as exhibited in Fig. 3f-h, where it is observed that a significant percentage of points (27%) fails to yield a valid solution in SHO analysis, as highlighted by white pixels in the mappings. Such a problem also casts doubts on the points wherein SHO analysis works, and indeed, mappings of intrinsic amplitude, resonant frequency, and quality factors all show not so subtle difference between SE- and DART-PFM, especially in rough regions. This highlights the advantage of SE over DART, which measures responses at only two excitation frequencies across resonance, as schematically shown in Fig. 1, and uses the difference between these two responses as feedback for resonance tracking. Such strategy often runs into difficulties: if the separation between two excitation frequencies is too small, they will easily fall out of resonance range during scanning and thus fail to track resonance shift; and if the separation is too large, then the responses are weak and the signal-to-noise ratios are low. For materials exhibiting substantial heterogeneity at the nanoscale, for instance near the grain boundaries [14,47,48] wherein the contact resonance frequency can shift significantly over a relatively short distance, resonance tracking of DART and thus SHO analysis often fail [15]. This is demonstrated on a PZT sample (Fig. 3f-h) that has a strong piezoelectric response yet a rough topography (Fig. 3b), which is not uncommon in practice, and the excitation frequency must shift substantially during scanning (Fig. 3g and Fig. S2). An excellent ferroelectric material such as PZT still suffers from such difficulty, and the issue is only more serious for other materials with weaker electromechanical coupling. Mappings acquired from SE-PFM, on the other hand, are free of such problems.

*Principal component analysis and its physical interpretation*

While SHO fitting works well under SE-PFM, it is a computationally an expensive process, taking an Intel Xeon E5-2695 CPU approximately 0.06s for one pixel and thus 1.09hr for a 256×256 mapping (or 5.8min for a parallel pool with 28 CPU workers). We thus resort to multivariate statistical tools such as PCA to speed up the analysis [16], as a sequence of images has been obtained under different excitation frequencies, ideal for PCA. Through orthogonal transformation, PCA converts a set of possibly correlated variables, in this case SE-PFM mappings under different excitation frequencies, to a set of linearly uncorrelated variable known as principal components. As a powerful unsupervised data analytic tool, PCA is widely used to compress and visualize multi-dimensional dataset, though its physical meaning is often unclear.

Here, we demonstrate that PCA not only speeds up our computation by four orders of magnitude, but also allows a clear physical interpretation of its modes in combination with SHO analysis. To this end, we recast 3D dataset of $A(\omega, x, y)$ into 2D dataset of $A(\omega, x)$, where 2D spatial grid is collapsed into 1D. This reshaped dataset can be viewed as a 2D matrix $\mathbf{A}$, such that each row of $\mathbf{A}$ represents a spatial mapping at a particular excitation frequency, while each column represents a spectrum of data spanning all excitation frequencies for a particular grid point. The details of following derivation are presented in the SI, wherein principal components of $\mathbf{A}$, *i.e.*, spatial eigenvectors $\mathbf{w}_i$ of the covariance matrix $\mathbf{A}^T\mathbf{A}$, are evaluated by singular value decomposition (SVD) [49]. Therefore, the $j^{th}$ row of $\mathbf{A}$, $\mathbf{A}_{j\text{-}row}$, can be regarded as a linear combination of $\mathbf{w}_i$ with coefficients $\xi_i$,

$$\mathbf{A}_{j-row} = \sum_i \xi_i(\omega_j) \cdot \mathbf{w}_i. \tag{2}$$

On the other hand, $\mathbf{A}_{j\text{-}row}$ can also be reformulated from Eq. (1) of SHO as detailed in SI,

$$\mathbf{A}_{j\text{-}row} = \sum_i \lambda_i(\omega_j) \cdot \boldsymbol{\alpha_i} = \sum_i \zeta_i(\omega_j) \cdot \boldsymbol{\beta_i}, \qquad (3)$$

where components $\{\boldsymbol{\alpha_i}\} = \mathbf{A_0 Q \omega_0} \circ [\mathbf{1}, \mathbf{Q}-\bar{\mathbf{Q}}, \boldsymbol{\omega_0}-\bar{\boldsymbol{\omega}}_0, (\boldsymbol{\omega_0}-\bar{\boldsymbol{\omega}}_0)^2, ...]$ inheriting all spatial variance of vectors $\mathbf{A_0}$, $\mathbf{Q}$, and $\boldsymbol{\omega_0}$ that are reshaped from intrinsic parameter mappings $A_0(x,y)$, $Q(x,y)$, and $\omega_0(x,y)$, $\mathbf{1}$ is a 1D vector all elements of which are 1, and $\bar{\mathbf{Q}} = \mathbf{1} \cdot \bar{Q}$ and $\bar{\boldsymbol{\omega}}_0 = \mathbf{1} \cdot \bar{\omega}_0$. Here, operator $\circ$ denotes the Hadamard product of two vectors and $\mathbf{A_0 Q \omega_0} = \mathbf{A_0} \circ \mathbf{Q} \circ \boldsymbol{\omega_0}$. As such, $\boldsymbol{\alpha}_1$ corresponds to $\mathbf{A_0 Q \omega_0}$ because it is always leading term in the 2D Taylor series, while the sequence following $\boldsymbol{\alpha}_i$ depends on the relative variation of $\mathbf{Q}$ and $\boldsymbol{\omega_0}$. To ensure the comparison with Eq. (2), we transform $\{\boldsymbol{\alpha_i}\}$ into a set of orthonormal basis $\{\boldsymbol{\beta_i}\}$ via Gram–Schmidt process [50], where $\boldsymbol{\beta}_1 = \boldsymbol{\alpha}_1 = \mathbf{A_0 Q \omega_0}$, $\boldsymbol{\beta}_2 = \boldsymbol{\alpha}_2 - \frac{\boldsymbol{\alpha}_2 \cdot \boldsymbol{\beta}_1}{\|\boldsymbol{\beta}_1\|^2}\boldsymbol{\beta}_1$, and $\boldsymbol{\beta}_3 = \boldsymbol{\alpha}_3 - \frac{\boldsymbol{\alpha}_3 \cdot \boldsymbol{\beta}_1}{\|\boldsymbol{\beta}_1\|^2}\boldsymbol{\beta}_1 - \frac{\boldsymbol{\alpha}_3 \cdot \boldsymbol{\beta}_2}{\|\boldsymbol{\beta}_2\|^2}\boldsymbol{\beta}_2$, after which $\{\boldsymbol{\beta_i}\}$ is normalized. The analogy between Eq. (3) and Eq. (2) is evident, suggesting that PCA components $\{\mathbf{w}_i\}$ corresponding to orthonormal basis $\{\boldsymbol{\beta_i}\}$ derived from SHO, which has a clear physical interpretation! In a completely parallel manner, the correspondence between PCA spectral eigenvectors and SHO expansion can be established by switching the row and column of $\mathbf{A}$, as detailed in the SI, *i.e.* between PCA spectral eigenvectors $\mathbf{Aw}_i$ and SHO spectral basis $\mathbf{A}\boldsymbol{\beta}_i$. In particular, elements of $\mathbf{Aw}_i$ represent the weight $\xi_i$ that $\mathbf{w}_i$ takes up in each scan according to Eq. (2).

To demonstrate this analysis, we compare the first three spectral eigenvectors of PCA versus the SHO expansion in Fig. 4a, using SE-PFM data presented in Fig. 2. Good agreement between PCA spectral eigenvectors $\mathbf{Aw}_i$ and the SHO spectral basis $\mathbf{A}\boldsymbol{\beta}_i$ is observed. The first three spatial eigenvectors of PCA are shown in Fig. 4b, in comparison with the first three SHO spatial basis in Fig. 4c, wherein good agreement is again observed, with the first spatial eigenvector maps $\boldsymbol{\beta}_1 = \mathbf{A_0 Q \omega_0}$, and second and third spatial eigenvectors maps $\boldsymbol{\beta}_2$ and $\boldsymbol{\beta}_3$, respectively, derived from $\boldsymbol{\alpha}_2 = \mathbf{A_0 Q \omega_0} \circ (\boldsymbol{\omega_0} - \bar{\boldsymbol{\omega}}_0)$ and $\boldsymbol{\alpha}_3 = \mathbf{A_0 Q \omega_0} \circ (\boldsymbol{\omega_0} - \bar{\boldsymbol{\omega}}_0)^2$ with Gram–Schmidt process. The structural similarity (SSIM) [51] for the first three pairs are evaluated to be 99.7%, 98.5%, and 95.5%, while the Pearson correlation coefficients (PCC) [52] are 90.06%, 91.15%, and 72.97%, as detailed in SI, confirming our analysis numerically.

Intuitively, the set of SE-PFM mappings under different frequencies contain two important information, the variation of the amplitude with respect to the spatial locations and with respect to excitation frequencies, which are interconnected in the original mappings of Fig. 2. Under PCA, the data are transformed, such that the spatial variation is best represented by the spatial eigenvectors, and the frequency variation is reflected in the spectral eigenvectors for each PCA mode. Note that the principal components are sorted by their eigenvalues in a descending manner, with the first principal component accounting for the maximum possible variability in the data, as shown by the scree plot of variance in Fig. S3. The physical interpretation of PCA eigenvectors, however, are often unclear, which we resolved in this work with the assistance of SHO analysis. Note that PCA takes only 0.24s for an Intel Xeon E5-2695 CPU to complete, which is four orders of magnitude faster than brute force fitting.

The spatial variation of intrinsic amplitude, resonant frequency, and quality factor, key material parameters of interests in this analysis, are not known in advance. Thus, in order to unambiguously establish our physical interpretation of PCA, we construct a model three-phase system numerically with given distribution of intrinsic amplitude, resonant frequency, and quality factor, as shown in Fig. 5a, from which corresponding SE-PFM mappings can be computed using SHO model followed by PCA analysis. Taylor expansion of SHO can then be carried out. The comparison of the first three spectral eigenvectors are shown in Fig. 5b, which agree with each other well. Meanwhile, the comparison of spatial eigenvectors for PCA (Fig. 5c) and SHO expansion (Fig. 5d) reveals a structural similarity of over 99.9% and a Pearson correlation coefficient of over 99.8% for first three modes, suggesting that the first three PCA spatial eigenvectors are $\boldsymbol{\beta}_1$, $\boldsymbol{\beta}_2$ and $\boldsymbol{\beta}_3$ derived from $(\boldsymbol{\alpha}_1, \boldsymbol{\alpha}_2, \boldsymbol{\alpha}_3) = \mathbf{A}_0 \mathbf{Q} \boldsymbol{\omega}_0 \circ [\mathbf{1}, \mathbf{Q} - \bar{\mathbf{Q}}, \boldsymbol{\omega}_0 - \bar{\boldsymbol{\omega}}_0]$, respectively. This set of studies thus confirm the physical interpretation of PCA modes, which can be used to substantially speed up the analysis.

**Conclusions**

Principal component analysis has been widely used to compress and visualize multi-dimensional dataset, though its physical meaning is often unclear. Scanning probe microscopy (SPM) measures a wide range of properties of sample through cantilever-sample interactions in term of cantilever dynamics, though intrinsic response is rather challenging to determine quantitatively, often interference by various cross-talks. Sequential excitation (SE) technique we developed, in combination with dynamics-based SHO model and data analytic PCA, allows us to overcome such difficulties through deep data methodology. Of particular interest is a clear mechanistic understanding of the unsupervised PCA acquired through sound physical principle, making it possible to speed up physical analysis by at least four orders of magnitude. The method can be easily implemented in any standard AFM without the need for any additional hardware and instrumentation. While the technique is demonstrated in terms of electromechanical coupling via PFM, the dynamics involved is universal in SPM, making the method widely applicable.

**Conflicts of interest**

There are no conflicts to declare.

**Acknowledgements**

We acknowledge National Key Research and Development Program of China (2016YFA0201001), US National Science Foundation (CBET-1435968), National Natural Science Foundation of China (11627801, 11472236), The Leading Talents of Guangdong Province (2016LJ06C372), and Shenzhen Knowledge Innovation Committee (KQJSCX20170331162214306, JCYJ20170818163902553). This material is based in part upon work supported by the State of Washington through the University of Washington Clean Energy Institute.

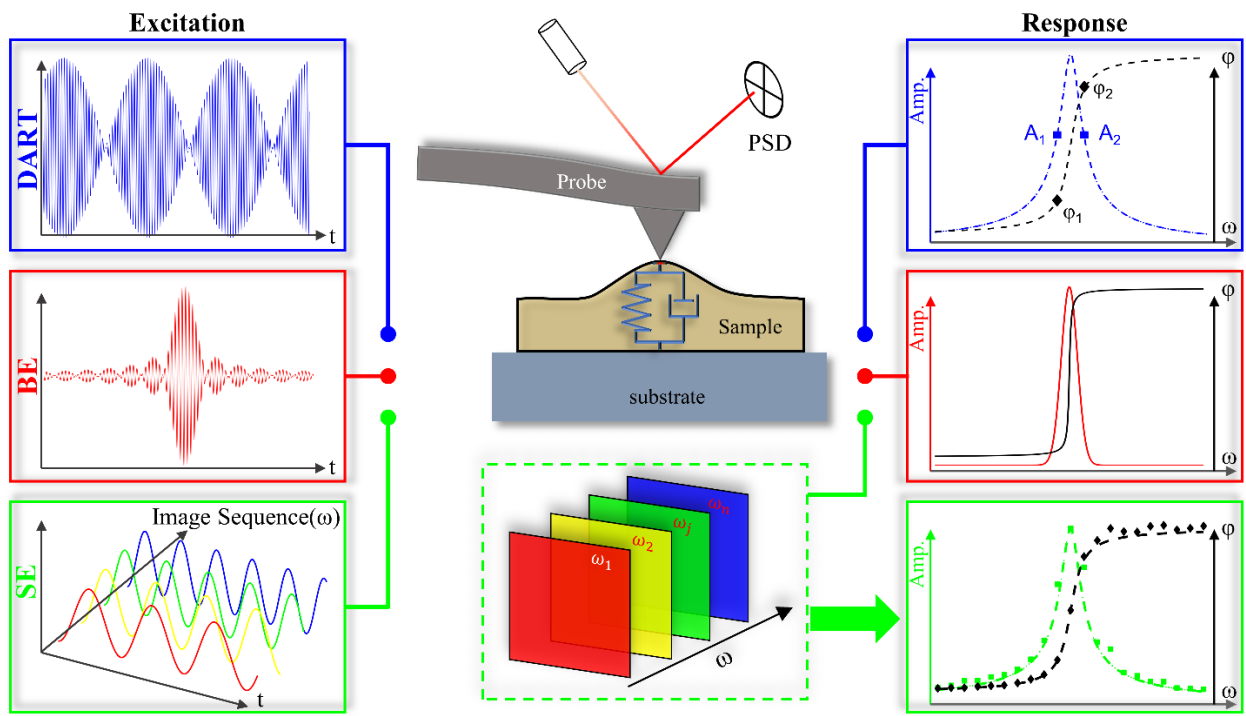

**Fig. 1** The schematics of dynamic SPM experiments based on DART, BE, and SE techniques, wherein AC waveform combining two distinct or a band of frequencies are synthesized to excite the sample under DART or BE, respectively, while a sequence of AC waveforms with different frequencies are used to excite the sample under SE.

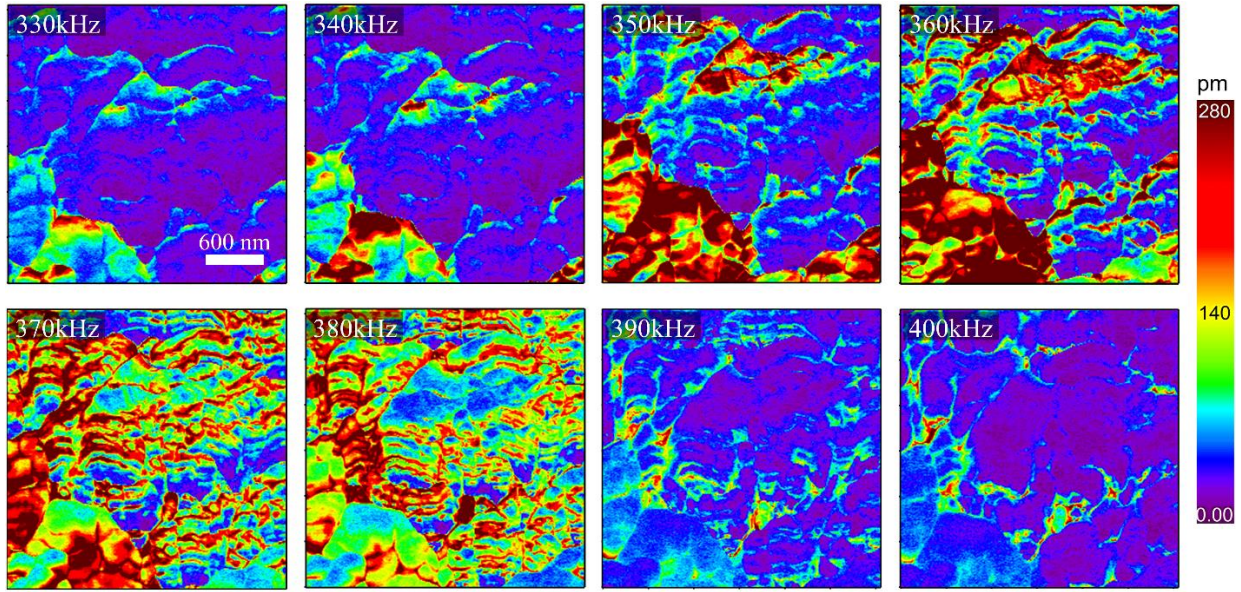

**Fig. 2** A sequence of SE-PFM amplitude mappings obtained at distinct frequencies in PZT ceramic.

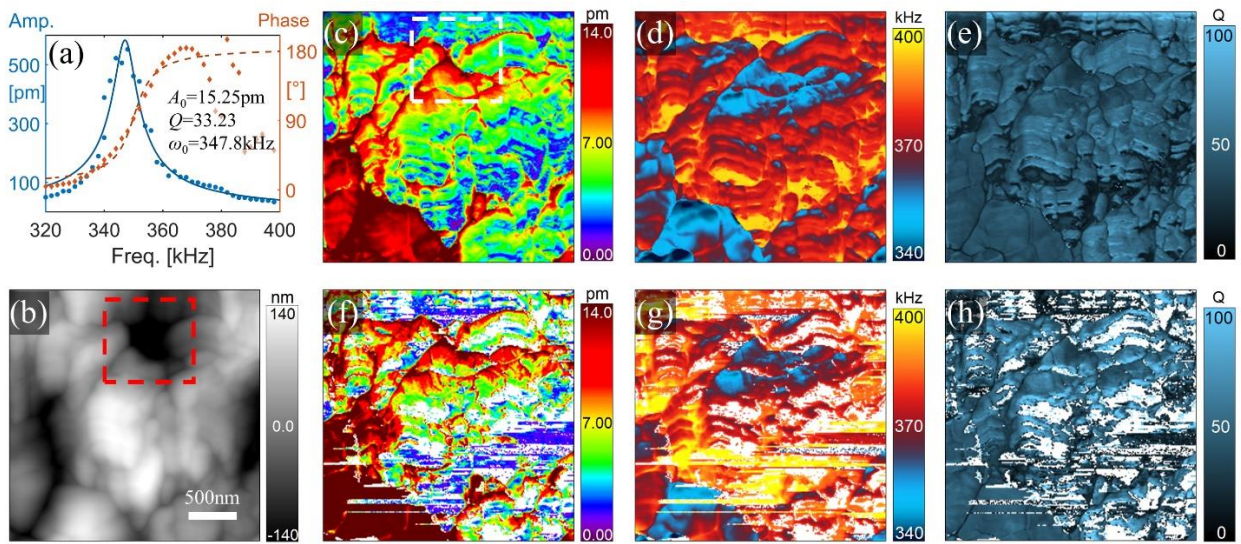

**Fig. 3** Comparison of PZT mappings acquired by SE-PFM and DART-PFM processed via SHO; (a) SHO fitting of SE-PFM spectrum data for one representative pixel; (b) rough topography mapping; and (c-e) reconstructed SE-PFM mappings of (c) intrinsic amplitude $A_0$, (d) resonance frequency $\omega_0$, and (e) quality factor $Q$; (f-h) reconstructed DART-PFM mappings of (f) the intrinsic amplitude, (g) resonance frequency, and (h) quality factor obtained, wherein white areas show point where SHO analysis fails.

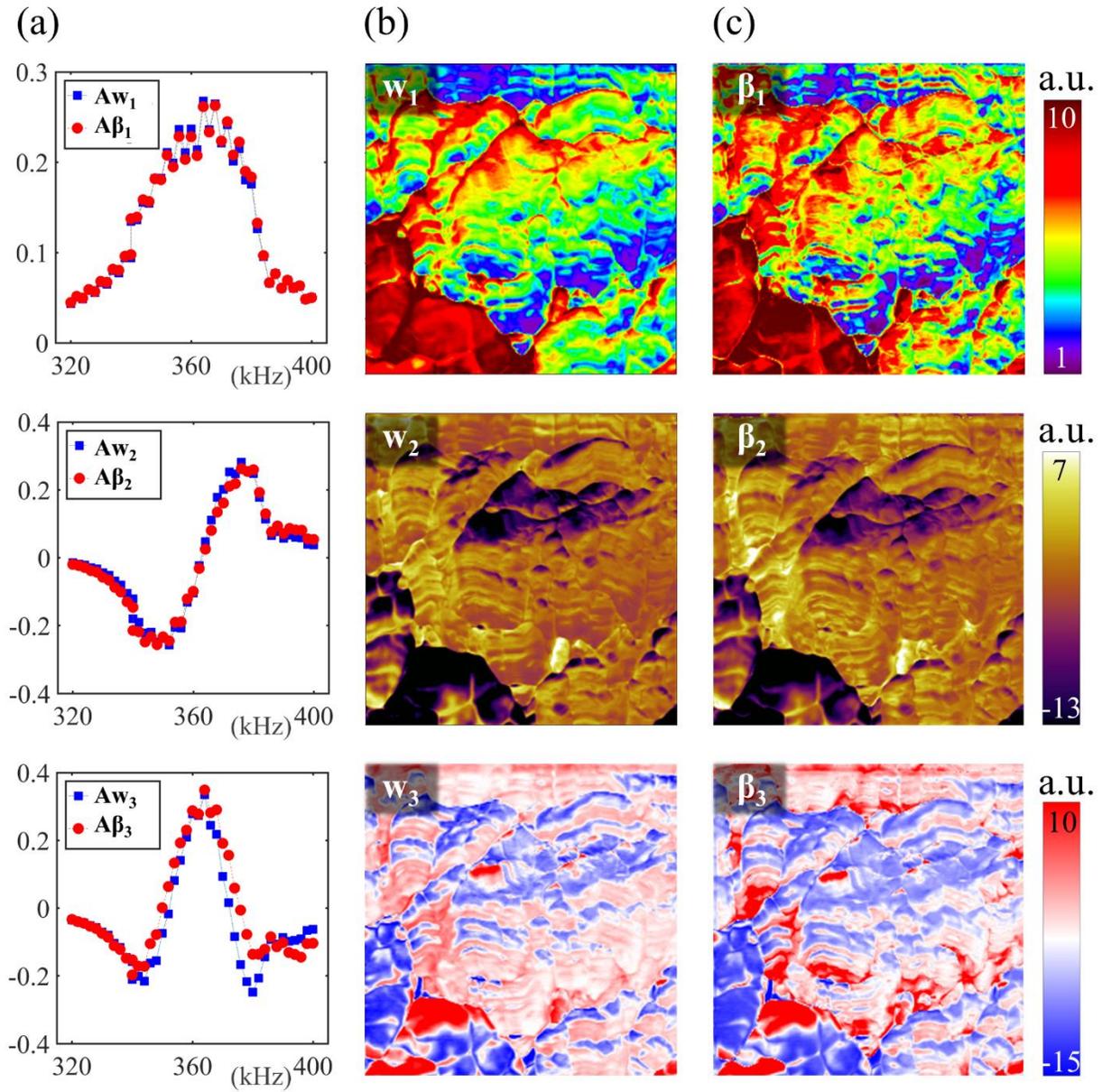

**Fig. 4** Comparison of PCA and SHO expansion for SE-PFM data of PZT; (a) first three PCA spectral eigenvectors in comparison with corresponding SHO spectral basis; (b) first three PCA spatial eigenvectors; (c) corresponding SHO spatial basis.

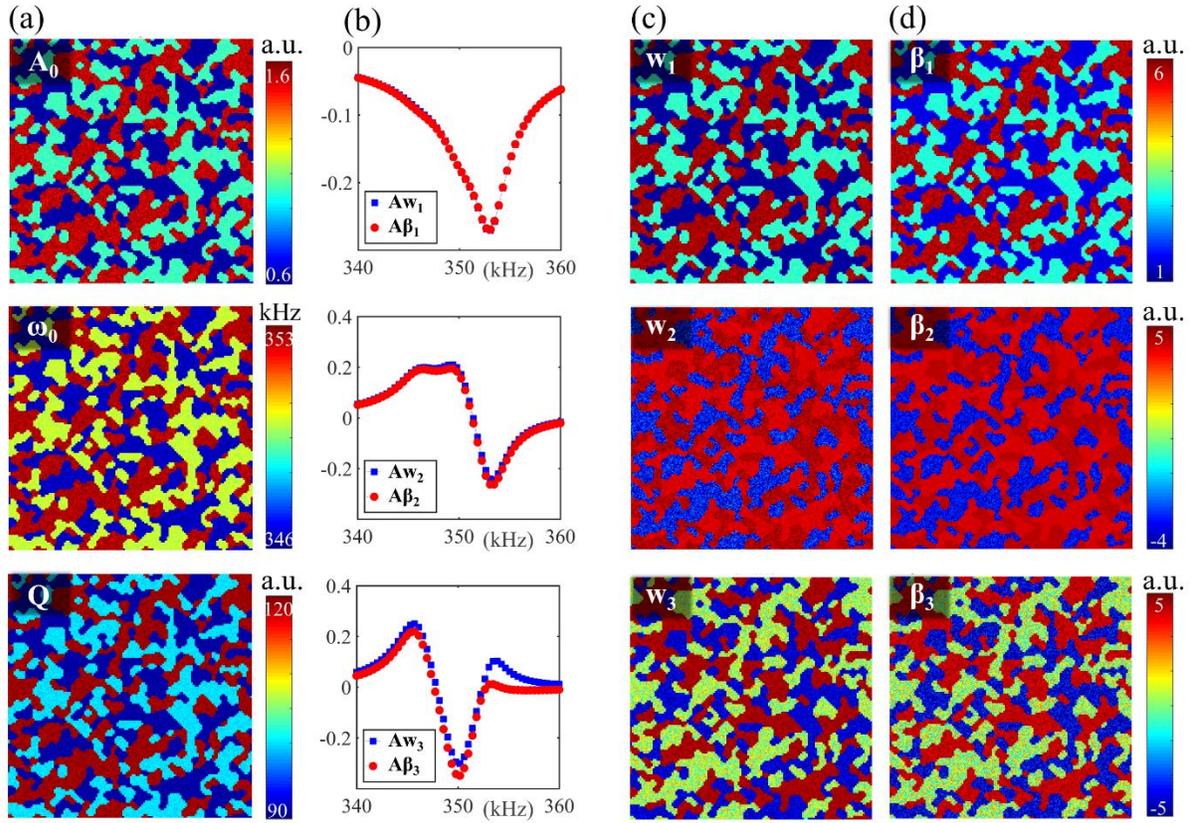

**Fig. 5** Comparison of PCA and SHO expansion for a three-phase model system with distributions of intrinsic amplitude, resonant frequency and quality factor specified in (a), from which the SE-PFM mappings can be constructed based on SHO; (b) comparison of first three spectral eigenvectors of PCA and corresponding SHO spectral basis; (c) the first three spatial eigenvectors of PCA; (d) corresponding SHO spatial basis.

**Mapping Intrinsic Electromechanical Responses at the Nanoscale via Sequential Excitation Scanning Probe Microscopy Empowered by Deep Data**

**Supporting Information**

**Methods**

*DART PFM*

An Asylum Research (AR) Cypher AFM was used to perform DART PFM measurements. An AC bias with an amplitude of 3V near the sample–probe resonance frequency was applied to the PZT disk to amplify the piezoresponse via a Nanosensors PPP-EFM with a spring constant of 2.6Nm$^{-1}$. The line scan rate was set to 0.57Hz to ensure the DART technique work smoothly.

*SE-PFM*

A series of single frequency PFM mappings were acquired under an AC voltage of 3V and excitation frequency ranging from 320 to 400 kHz with a 2kHz increment. The scan rate was set as 1.5Hz. Therefore, a total of 41 PFM mappings were obtained by using the programmable function Macrobuilder of AR software. All raw amplitude data were then imported into MATLAB as 41 256×256 matrices. Subpixel image registration were performed for all images by cross-correlation to correct for inevitable topography drifts during the experiment.

*Image registration*

We take the deflection mapping of the first scan as a benchmark to do the image registration for other scans. The obtained translation is applied to the amplitude and phase mappings as well to correct drifting. An open-source algorithm is adopted in this process, which obtains an initial translation estimation of the cross-correlation peak and refines it by a matrix-multiply discrete Fourier transform (DFT) [1]. **Mov. S1** compares the original deflection map with the registered mappings of deflection, amplitude and phase, where each frame represents data acquired under a specific frequency.

*SHO Fitting and $R^2$ Mapping*

Default Fit function with nonlinear least-squares method in MATLAB is used to perform SHO fitting, which returns the mappings of $A_0$, $\omega_0$, $Q$, and $R^2$. $R^2$ is a statistical measure of how close the data are to the fitted regression line. It ranges from 0 to 1. $R^2=1$ indicates that the fitted data $f_i$ explains all variability in observed data $y_i$, while $R^2=0$ indicates no 'linear' relationship between them. The most general definition of $R^2$ is

$$R^2 = 1 - \frac{SS_{res}}{SS_{tot}},$$

where $SS_{tot} = \sum_{i=1}^{n}(y_i - \bar{y})^2$, $SS_{res} = \sum_{i=1}^{n}(y_i - f_i)^2$.

*PCA Analysis*

PCA function based on SVD in MATLAB is used to perform this analysis. The variance data plotted in Fig.S3, which are also known as the eigenvalues of the covariance matrix, are from an output vector "latent" of PCA function.

*Structural similarity*

SSIM is an index for measuring the similarity between two images, which is defined as the mean of the local SSIM value map $SSIM(x,y)$:

$$SSIM(x,y) = \frac{(2\mu_a\mu_b + c_1)(2\sigma_{ab} + c_2)}{(\mu_a^2 + \mu_b^2 + c_1)(\sigma_a^2 + \sigma_b^2 + c_2)},$$

where $\mu_a$, $\mu_b$, $\sigma_a$, $\sigma_b$, $\sigma_{ab}$ are the average, variance, and covariance of 4×4 windows a and b that are centered in the pixel (x, y) of two images. $c_1$, $c_2$ are two variables to stabilize the division with weak denominator.

*Pearson correlation coefficient*

PCC is a measurement of the linear correlation between two vectors X and Y, which are reshaped from two images, respectively. It ranges from +1 and −1, where 1 means total positive linear correlation, 0 means no linear correlation, and −1 means total negative linear correlation,

$$P_{X,Y} = \frac{\sigma_{XY}}{\varsigma_X \varsigma_Y}$$

where $\sigma_{XY}$ is the covariance and $\varsigma_X$, $\varsigma_Y$ are the standard deviation of X and Y.

**Derivation of PCA and SHO on spatial mode**

*PCA modes*

Under SVD, we have $A = U\Sigma W^T$, where $\Sigma$ is a diagonal matrix of positive singular values $\sigma_i$. The columns of $U$ and $W$ are left- and right-singular vectors $\mathbf{u}_i$ and $\mathbf{w}_i$, respectively. Since unitary matrices $U$ and $W$ satisfy $U^T U = W^T W = I$, we have

$$(A^T A) \cdot W = W\Sigma^T U^T U\Sigma W^T \cdot W = W \cdot \Sigma^2$$
$$\Rightarrow (A^T A) \cdot \mathbf{w}_i = \sigma_i^2 \cdot \mathbf{w}_i \quad , \quad (S1)$$

where $\mathbf{w}_i$ are defined as PCA spatial eigenvectors sorted in the order of decreasing $\sigma_i^2$.

*SHO modes*

For any given pixel scanned at a specific frequency $\omega_j$, $A(\omega_j)$ can be reformulated as follows by plugging Eq. (1.2) into Eq. (1.1),

$$A(\omega_j) = \frac{A_0 \omega_0^2}{\sqrt{[\frac{1}{\tan\phi(\omega_j, Q, \omega_0)}]^2 + 1}} \frac{Q}{\omega_0 \omega_j}$$

$$= \sin\phi(\omega_j, Q, \omega_0) \cdot A_0 \omega_0^2 \frac{Q}{\omega_0 \omega_j} \quad (S2)$$

$$= \eta(\omega_j, Q, \omega_0) \cdot A_0 Q \omega_0$$

where $\eta(\omega_j, Q, \omega_0) = \sin\phi(\omega_j, Q, \omega_0)/\omega_j$. Then it can be expanded into 2D Taylor series around the spatial average $(\bar{Q}, \bar{\omega}_0)$ of the whole scanned area as,

$$A(\omega_j) = A_0 Q \omega_0 \cdot \eta(\omega_j, Q, \omega_0)$$
$$= A_0 Q \omega_0 \cdot [\eta_0(\omega_j, \bar{Q}, \bar{\omega}_0) + \frac{\partial \eta}{\partial Q}\bigg|_{(\omega_j, \bar{Q}, \bar{\omega}_0)} \cdot \Delta Q + \frac{\partial \eta}{\partial \omega_0}\bigg|_{(\omega_j, \bar{Q}, \bar{\omega}_0)} \cdot \Delta\omega_0 + \frac{\partial^2 \eta}{\partial^2 \omega_0}\bigg|_{(\omega_j, \bar{Q}, \bar{\omega}_0)} \cdot (\Delta\omega_0)^2 + ...]$$
$$= A_0 Q \omega_0 \cdot [\lambda_1(\omega_j) + \lambda_2(\omega_j) \cdot (Q - \bar{Q}) + \lambda_3(\omega_j) \cdot (\omega_0 - \bar{\omega}_0) + \lambda_4(\omega_j) \cdot (\omega_0 - \bar{\omega}_0)^2 + ...]$$

... (S3)

where $\lambda_i(\omega_j)$ are coefficients that only depends on $\omega_j$ since $(\bar{Q}, \bar{\omega}_0)$ are constant. This expansion can be further generalized to the whole map as,

$$\mathbf{A}_{j-row} = \mathbf{A_0 Q \omega_0} \circ [\lambda_1(\omega_j) + \lambda_2(\omega_j) \cdot (\mathbf{Q} - \bar{\mathbf{Q}}) + \lambda_3(\omega_j) \cdot (\boldsymbol{\omega_0} - \bar{\boldsymbol{\omega}}_0) + \lambda_4(\omega_j) \cdot (\boldsymbol{\omega_0} - \bar{\boldsymbol{\omega}}_0)^2 + ...]$$
$$= \sum_i \lambda_i(\omega_j) \cdot \boldsymbol{\alpha_i}$$

... (S4)

where $\{\boldsymbol{\alpha_i}\} = \mathbf{A_0 Q \omega_0} \circ [\mathbf{1}, \mathbf{Q} - \bar{\mathbf{Q}}, \boldsymbol{\omega_0} - \bar{\boldsymbol{\omega}}_0, (\boldsymbol{\omega_0} - \bar{\boldsymbol{\omega}}_0)^2, ...]$ and $\mathbf{A_0 Q \omega_0} = \mathbf{A_0} \circ \mathbf{Q} \circ \boldsymbol{\omega_0}$ is the Hadamard product of 1D vectors $\mathbf{A_0}$, $\mathbf{Q}$, and $\boldsymbol{\omega_0}$ that are reshaped from intrinsic parameter mappings $A_0(x,y)$, $Q(x,y)$, and $\omega_0(x,y)$, respectively. $\mathbf{1}$ is a 1D vector, all elements of which are 1, while $\bar{\mathbf{Q}} = \mathbf{1} \cdot \bar{Q}$ and $\bar{\boldsymbol{\omega}}_0 = \mathbf{1} \cdot \bar{\omega}_0$.

**Derivation of PCA and SHO on frequency mode**

*PCA modes*

After switching the row and column of $\mathbf{A}$, we get $\mathbf{A}^T$, where now each row represents a spectrum of data spanning all excitation frequencies for a particular grid point. Consequently, the PCA of $\mathbf{A}^T$ gives principal spectral modes by computing the eigenvectors of the new covariance matrix $\mathbf{AA}^T$. A simple way to do this is left multiplying $\mathbf{A}$ on the both sides of (Eq. S1),

$$\mathbf{A} \cdot (\mathbf{A}^T \mathbf{A}) \cdot \mathbf{w}_i = \mathbf{A} \cdot \sigma_i^2 \cdot \mathbf{w}_i \Rightarrow (\mathbf{AA}^T) \cdot \mathbf{A}\mathbf{w}_i = \sigma_i^2 \cdot \mathbf{A}\mathbf{w}_i, \qquad (S3)$$

so PCA spectral eigenvectors can be generated by normalizing $\{\mathbf{A}\mathbf{w}_i\}$. According to Eq. (2) and the orthogonality of $\{\mathbf{w}_i\}$, we have $\mathbf{A}\mathbf{w}_i = \boldsymbol{\xi}_i$, which is a vector representing the weight that $\mathbf{w}_i$ takes up in each scan.

*SHO modes*

Since $\{\boldsymbol{\beta}_i\}$ resembles $\{\mathbf{w}_i\}$, it is expected that $\{\mathbf{A}\boldsymbol{\beta}_i\}$ should be close to $\{\mathbf{A}\mathbf{w}_i\}$ as well. Besides, $\mathbf{A}\boldsymbol{\beta}_i = \boldsymbol{\zeta}_i$ also represents a vector weight that $\boldsymbol{\beta}_i$ takes up in each scan based on Eq. (3).

**Additional Data**

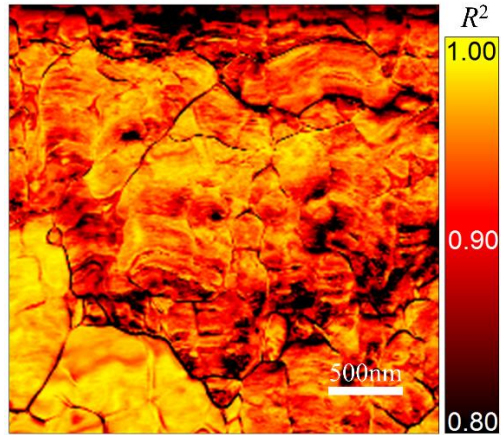

**Fig. S1** Mapping of $R^2$

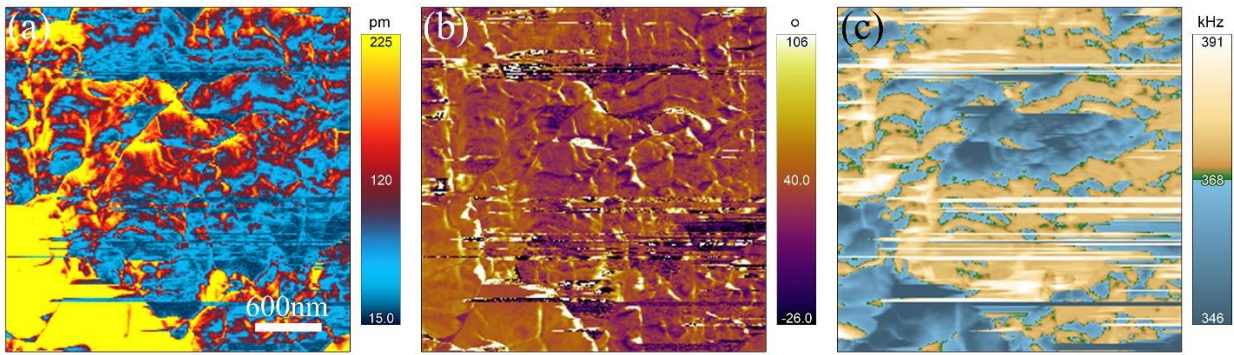

**Fig. S2** DART-PFM PZT mappings of amplitude (a) and phase (b) acquired at lower frequency of two excitation frequency (c).

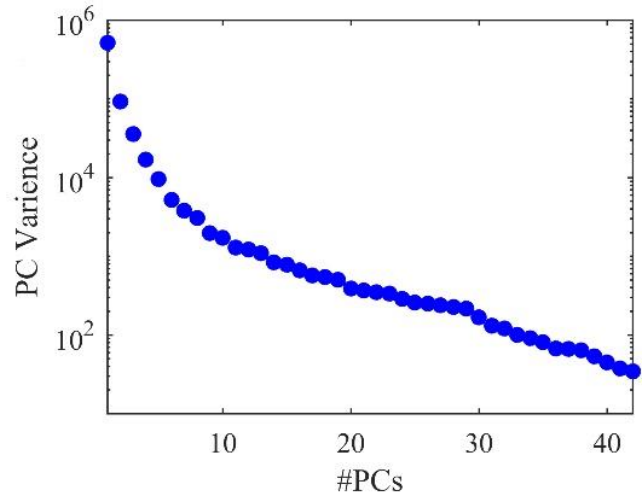

**Fig. S3** Scree plot of PCA variance for PZT.

https://drive.google.com/file/d/1C3_drUCIRBXWmICOK4vYCM4Kg31OYUyv/view

**Mov. S1** Realignment of SE-PFM data set to correct for drifting among different scans.